\documentclass{iopjournal}
\usepackage[T1]{fontenc}
\usepackage{amsmath}
\usepackage{amssymb}
\usepackage{ragged2e}

\begin{document}
\justifying

\articletype{Paper} 

\title{Binary Neutron Star Merger Simulations with Microphysical Equation of State using Spritz}

\author{Fatemeh Hossein Nouri$^1$\orcid{0000-0003-1449-0824}, Jay V. Kalinani$^2$\orcid{0000-0002-2945-1142}, Bruno Giacomazzo$^{3,1}$\orcid{0000-0002-6947-4023}, Riccardo Ciolfi$^{4,5}$\orcid{0000-0003-3140-8933}, and Albino Perego$^{6,7}$\orcid{0000-0002-0936-8237}}

\affil{$^1$INFN, Sezione di Milano-Bicocca, Piazza della Scienza 3, I-20126 Milano, Italy}

\affil{$^2$The Grainger College of Engineering, Department of Physics \& Illinois Center for Advanced Studies of the Universe, University of Illinois Urbana-Champaign, Urbana, Illinois 61801, USA}

\affil{$^3$Dipartimento di Fisica G. Occhialini, Universit\`a di Milano-Bicocca, Piazza della Scienza 3, I-20126 Milano, Italy}

\affil{$^4$INAF, Osservatorio Astronomico di Padova, Vicolo dell’Osservatorio 5, I-35122 Padova, Italy}

\affil{$^5$INFN, Sezione di Padova, Via Francesco Marzolo 8, I-35131 Padova, Italy}

\affil{$^6$Dipartimento di Fisica, Universit\`a di Trento, via Sommarive 14, 38123 Trento, Italy}

\affil{$^7$INFN-TIFPA, Trento Institute for Fundamental Physics and Applications, via Sommarive 14, 38123 Trento, Italy}


\email{fatemeh.hosseinnouri@unimib.it}


\begin{abstract}
\justifying

\noindent We present the first binary neutron star merger simulations performed with the general-relativistic magnetohydrodynamics (GRMHD) \textsc{Spritz} code employing a finite-temperature tabulated equation of state. Two magnetized binaries with the LS220 equation of state and a GW170817-like chirp mass, differing only in their mass ratios ($q=1$ and $q=0.7$), are evolved from the late inspiral through merger, hypermassive neutron-star (HMNS) formation, and delayed collapse to a black hole (BH). We investigate how the binary mass ratio influences the postmerger evolution, jet-launching conditions, and matter ejection. 
We find that the equal-mass binary forms a slightly longer-lived HMNS, allowing more efficient magnetic-field amplification during the post-merger evolution. As a result, it accumulates larger magnetic fluxes, develops a magnetically dominated funnel, and launches a collimated magnetically driven polar outflow.
In contrast, the unequal-mass binary ejects approximately 50\% more mass but exhibits weaker magnetic-field amplification, with the polar region remaining turbulent and baryon loaded throughout our simulation, suggesting that the development of a magnetically dominated funnel is likely delayed.

\end{abstract}

\section{Introduction}

Binary neutron-star (BNS) mergers are among the most energetic events in the Universe. They produce gravitational waves (GWs), gamma-ray bursts (GRBs), and kilonovae powered by the radioactive decay of r-process nuclei. The first observed BNS merger, GW170817, established the connection between all three messengers by combining GW, gamma-ray, and optical/infrared observations in a single event~\cite{Abbott-2017a,Abbott-2017b,Abbott-2017c,Goldstein-2017,Tanvir-2017,Pian-2017}. Interpreting these observations requires accurate numerical simulations that incorporate realistic nuclear equations of state (EOS), neutrino physics, and magnetic fields. The EOS determines the thermodynamic properties of matter at supra-nuclear densities and influences the GW signal, the properties of the ejecta, and the fate of the merger remnant~(see e.g. \cite{Baiotti-2017,Radice-2020b} for recent reviews and references therein).
Magnetic fields, on the other hand, can be amplified through different mechanisms, including magnetorotational instability (MRI)~\cite{Balbus-1991} and the Kelvin-Helmholtz instability (KHI)~\cite{Rasio-1999}, and govern angular-momentum transport, drive jet launching, and contribute to mass ejection through magnetically driven turbulence.

Early 3D GRMHD simulations of magnetized BNS mergers (e.g.~\cite{Giacomazzo-2011,Rezzolla-2011}) showed that magnetic fields can be amplified by orders of magnitude due to the KHI and MRI during merger and postmerger; however, this amplification strongly depends on resolution.
Later, Kiuchi and collaborators performed ultra-high-resolution GRMHD simulations that more efficiently resolved magnetically-driven instabilities~\cite{Kiuchi-2014,Kiuchi-2015a,Kiuchi-2015b,Kiuchi-2018}. They found that magnetic fields can be amplified up to $\sim 10^{16}$ G, establishing the fact that magnetic amplification is extremely resolution-dependent.
However, the relevant small-scale dynamics cannot be fully resolved with the current feasible resolutions in GRMHD simulations. Therefore, effective models based on subgrid and Large-Eddy approaches are necessary to capture MHD turbulence and magnetic field amplification occurring at scales smaller than the employed grid spacing \cite{Giacomazzo-2015,Palenzuela-2015,Radice-2017,Radice-2020,Palenzuela-2022}.

Closely related to magnetically driven jets and outflows, several groups have investigated the conditions required for launching relativistic jets following binary neutron star mergers. Ruiz et al. (2016-2021) showed that black hole (BH)-disk systems formed after a delayed collapse can power jets depending on the initial configuration and amplification of the magnetic fields~\cite{Ruiz-2016,Ruiz-2019,Ruiz-2020,Ruiz-2021}. 
In contrast, for the prompt-collapse configurations considered in their study, they did not observe the formation of an incipient jet or a magnetically dominated polar funnel~\cite{Ruiz-2017,Ruiz-2021}. More recently, however, Hayashi et al. (2025)~\cite{Hayashi-2025} showed that an asymmetric prompt-collapse merger can still launch a relativistic jet if the merger leaves behind a sufficiently massive turbulent accretion disk that is evolved long enough ($\sim 1.5$\,s) to amplify and reorder the magnetic field.
Other GRMHD studies by Ciolfi et al. (2017, 2019)~\cite{Ciolfi-2017,Ciolfi-2019} confirmed that the lifetime of the remnant and the degree of magnetic amplification are key factors for launching relativistic jets and outflows.
Later, Ciolfi (2020) showed that a magnetically-driven collimated outflow is generated in a long simulation ($\sim 250$ ms) by a strongly magnetized long-lived remnant. However, the properties of such an outflow were found largely incompatible with short GRB jets, favoring the delayed BH-disk scenario~\cite{Ciolfi-2020}. 
More recently, Kiuchi et al. (2024)~\cite{Kiuchi-2024-alpha} demonstrated that, at sufficiently high resolution, MRI-driven turbulence can sustain an $\alpha \Omega$-dynamo that regenerates the large-scale poloidal magnetic field in a long-lived remnant, possibly enabling the formation of a relativistic jet on secular timescales without BH formation.

On the microphysics side, the community moved to include sophisticated finite-temperature, composition dependent EOS and neutrino transport. Sekiguchi et al. (2015, 2016)~\cite{Sekiguchi-2015,Sekiguchi-2016} and Palenzuela et al. (2015)~\cite{Palenzuela-2015} were among the first groups to systematically combine full GR merger dynamics, tabulated EOS, and neutrino leakage in 3D merger simulations — building on earlier Newtonian and relativistic works (Ruffert 1997; Rosswog 2003; Dessart 2009; Sekiguchi 2011; Korobkin 2012; Perego 2014)\cite{Ruffert:1996by,Rosswog-2003,Dessart-2009,Sekiguchi-2011,Korobkin-2012,Perego-2014}.
Kastaun et al. (2016)~\cite{Kastaun-2016} investigated the internal structure and thermal state of a BNS remnant with tabulated EOS that remains stable, providing a baseline for understanding different remnant fates.
Radice and collaborators (2017–2020) built extensively in this direction: they systematically explored EOS and neutrino effects on ejecta composition and kilonova signatures, using a general-relativistic large-eddy simulation (GRLES) approach to model unresolved magnetically-driven turbulence in merger remnants~\cite{Radice-2017,Radice-2018a,Radice-2018b,Radice-2020}. 
For long-term ejecta and nucleosynthesis targeted to the GW170817 event, Nedora et al. (2021)~\cite{Nedora-2021} ran long-term simulations with microphysics and turbulence modeling, identifying spiral-wave winds and showing how combined dynamical and wind ejecta can reproduce r-process abundance patterns under certain conditions. 

Despite these significant advances, relatively few GRMHD studies have combined realistic finite-temperature microphysics with self-consistent magnetic field evolution to investigate how the binary mass ratio influences the postmerger evolution in a delayed-collapse remnant scenario.
Motivated by this, we present new simulations of magnetized binary neutron star mergers performed with \textsc{Spritz}. Our goal is to investigate the impact of realistic microphysics and magnetic fields on the postmerger evolution and to estimate the role of the binary mass ratio in shaping the observable multi-messenger signatures. We consider two magnetized binary systems with the LS220 equation of state and a GW170817-like chirp mass, differing in their mass ratios ($q=1$ and $q=0.7$). We investigate the dynamical evolution of the merger remnants, the amplification of the magnetic field, and the resulting GW emission. We further explore the conditions for launching magnetically driven jets, including the development of collimated polar outflows during the HMNS phase and the formation of a magnetically dominated funnel that may provide favorable conditions for the activation of the Blandford-Znajek mechanism following BH formation. Finally, we examine the mass ejection from both systems and quantify how the binary mass ratio influences the amount, geometry, and evolution of the ejecta.

The remainder of this paper is organized as follows. In Sec.~\ref{sec:methods}, we describe the numerical framework, initial data, and simulation setup. In Sec.~\ref{sec:results}, we present the dynamical evolution and gravitational wave signals extracted from the merger and remnants. We also report on the evolution of the magnetic field, the conditions for the emergence of incipient jets, and the characteristics of the ejecta, placing our findings in the context of previous GRMHD studies. Finally, Sec.~\ref{sec:conclusions} summarizes the main results and discusses their broader implications for understanding the multi-messenger signatures of binary neutron-star mergers. The appendix is dedicated to a numerical resolution study and MRI analysis. 

\section{Methods}
\label{sec:methods}

\subsection{Numerical methods and simulation setups}

The simulations are performed with the \textsc{Spritz} code~\cite{Cipolletta:2020,Cipolletta:2021,Kalinani:2022,Spritz_v100}, a GRMHD framework built on the \textsc{Einstein Toolkit}~\cite{Loffler:2011ay,EinsteinToolkit2024} infrastructure. The GRMHD equations are solved in conservative form using a high-resolution shock-capturing scheme. Primitive variables are reconstructed at cell interfaces using the fifth-order WENOZ method~\cite{WENOZ}, while numerical fluxes are computed with the HLLE Riemann solver~\cite{HLLE}. 
The spacetime evolution is carried out within the CCZ4 formulation~\cite{Alic2012} of the Einstein equations implemented in the \textsc{MacLachlan} module of \textsc{EinsteinToolkit}, with damping parameters $\kappa_1 = 0.065$ and $\kappa_3 = 0.5$, chosen to control constraint violations and improve stability, particularly after BH formation~\cite{Alic2013}. 
Magnetic fields are evolved via a vector potential formulation, employing a generalized Lorenz gauge to enforce the divergence-free condition~\cite{Etienne2012Lorenz}. 
Neutrino effects are implemented through a leakage scheme based on the publicly available ZelmaniLeak module \cite{Ott2012}. 
However, in the present work, neutrino cooling is
disabled and the electron fraction $Y_e$ is simply advected throughout the evolution.
The numerical methods and their validation, including the coupling to tabulated equations of state and neutrino leakage, are described in Cipolletta et al. 2020~\cite{Cipolletta:2020}, Cipolletta et al. 2021~\cite{Cipolletta:2021}, and Kalinani et al. 2022~\cite{Kalinani:2022}. The microphysics framework has been successfully tested for equilibrium configurations of an isolated NS~\cite{Cipolletta:2021}. 
In this work, we employ an updated version of the code that includes corrections to previously identified numerical issues, together with several improvements aimed at enhancing the robustness of BNS merger simulations with tabulated EOSs, publicly available on \textsc{Zenodo}~\cite{Spritz_v111}. Close to BH formation, when the minimum lapse drops below $\alpha=0.22$, we restrict the rest-mass density to the upper bound of the EOS table in order to avoid failures associated with EOS-table interpolation. We also impose a ceiling on the magnetization $\sigma=b^2/\rho$ by increasing the density according to $\rho=b^2/\sigma_{\rm max}$ wherever $\sigma>\sigma_{\rm max}$. A default value of $\sigma_{\rm max}=1000$ is adopted, while lower magnetization thresholds are used for the $q=1$ simulations where required to maintain numerical stability.

We consider two configurations of magnetized BNS systems with the same chirp mass as GW170817: an equal-mass binary with component gravitational masses $M_1=M_2=1.3624\,M_\odot$, and an unequal-mass binary with $M_1=1.6363\,M_\odot$ and $M_2=1.1454\,M_\odot$, corresponding to $q=M_2/M_1=0.7$.
The initial data are generated using the \textsc{FUKA} code~\cite{Papenfort:2021}, imposing a coordinate separation of 45\,km between the two stars. We employ a resolution corresponding to a finest grid spacing of $dx = 0.18$ (in units of $M_\odot$), which corresponds to approximately $dx \simeq 266\,\mathrm{m}$. The computational domain extends to an outer boundary located at 3200\,km in all spatial directions and employs 8 refinement levels. 

In this work, we employ the LS220 EOS, a Skyrme-based finite-temperature, composition-dependent nuclear EOS based on the Lattimer–Swesty model with an incompressibility modulus of $K=220$ MeV~\cite{Lattimer1991}. 
The binaries are initialized as cold, beta-equilibrated configurations with a uniform temperature of $T=0.01~\mathrm{MeV}$.
This EOS provides thermodynamic quantities over a wide range of densities, temperatures, and electron fractions, making it well-suited for dynamical simulations of BNS mergers. Although LS220 does not incorporate chiral effective field theory constraints at high densities and provides a too large value for the slope of the symmetry energy at saturation density (see e.g. \cite{Tews:2016jhi}), it remains one of the most widely used EOS choices in the literature, which allows for comparison with a large number of existing studies. Furthermore, for the masses considered here, LS220 is expected to lead to a delayed collapse scenario, with BH formation occurring $\sim 15$–$35$ ms after merger. Such a configuration is particularly favorable for the development of conditions associated with relativistic jet formation. Finally, this choice enables us to demonstrate the capability of our numerical framework to consistently handle all phases of the evolution—from inspiral and merger to the formation and evolution of a HMNS remnant, and its eventual collapse to a BH surrounded by an accretion disk.

The initial magnetic field is confined to the interior of each neutron star and is constructed to be purely poloidal via a vector potential prescription. Specifically, we define the azimuthal component of the vector potential as $A_{\phi} \equiv A_b \varpi ^2 \max{(P - P_{cut}, 0)}^{n_s}$, where $\varpi$ is the cylindrical radius, $P$ is the fluid pressure, and $A_b$ is a normalization constant that sets the magnetic field strength. The cutoff pressure is chosen as $P_{\mathrm{cut}} = 0.04 P_{\max}$, where $P_{\max}$ is the initial maximum pressure. The parameter $n_s = 2$ controls the degree of differentiability of the magnetic field profile. The constant $A_b$ is chosen such that the initial maximum magnetic field strength reaches $B_{\max} \sim 2 \times 10^{16}\mathrm{G}$. This relatively large field strength is adopted to compensate for the limited numerical resolution, which cannot fully capture small-scale magnetic field amplification processes such as the KHI and MRI. This choice allows us to probe the dynamics of strongly magnetized postmerger remnants and to create more favorable conditions for the potential development of relativistic jets. 

\subsection{Post-analysis methods}

\subsubsection{PostCactus and extraction of GW}

Post-processing and data analysis are performed using the \textsc{PostCactus} Python package, a publicly available toolkit developed for the analysis and visualization of numerical relativity simulations~\cite{Kastaun2021PostCactus}. \textsc{PostCactus} provides routines for reading and processing \textsc{Einstein Toolkit} outputs, including GW extraction and spacetime diagnostics. In this work, the package is used to extract GW strains from the scalar $\Psi_4$, compute the corresponding GW spectra, and generate several of the visualizations presented throughout this paper.

\subsubsection{Total magnetic energy measurement}

To quantify the electromagnetic energy content of the system, we compute the total magnetic energy following the definition introduced by Duez et al. 2006~\cite{Duez2006}. The electromagnetic energy is evaluated as
\begin{equation}
E_{EM} = \int_V \frac{n_{\mu}n_{\nu}T^{\mu\nu}}{\alpha u^0}dV \, ,
\label{eq:E_em}
\end{equation}
where $T^{\mu\nu}_{\mathrm{EM}}$ is the electromagnetic contribution to the stress-energy tensor, $n^\mu$ is the timelike unit normal to the spatial hypersurface, $\alpha$ is the lapse function, and $u^0$ is the temporal component of the fluid four-velocity. The integration is performed over the computational volume $V$. This quantity provides a global measure of the electromagnetic energy stored in the system and allows us to track the amplification of magnetic energy throughout the inspiral, merger, and post-merger evolution of the remnant.

\subsubsection{Magnetic flux measurement:}

To quantify the magnetic field structure and its evolution during the post-merger phase, we compute the radial magnetic flux through spherical surfaces of constant coordinate radius. In particular, we evaluate the flux associated with the radial component of the magnetic field as
\begin{equation}
\left| \Phi_{B,r} \right| = \frac{1}{2} \int_{S(r)} \left| B^r \right| dA \, ,
\label{eq:Phi_Br}
\end{equation}
where $B^r$ is the radial component of magnetic field and $dA = r^2 \sin\theta~d\theta~d\phi$ is the surface element on a sphere of radius $r$. The factor of $1/2$ accounts for the integration over both hemispheres when using the absolute value of the radial field component. This diagnostic provides a useful measure of the large-scale poloidal magnetic field threading the polar region and is commonly employed to assess the conditions for magnetically driven outflows and potential jet formation in BNS merger remnants.

\subsubsection{Electromagnetic luminosity:}
\label{sec:lem_diagnostics}

To quantify the strength and angular structure of the magnetically-driven outflow, we analyze the Poynting luminosity extracted on spherical surfaces at different radii using the multipolar decomposition provided by the simulation outputs. The electromagnetic energy flux is decomposed in terms of scalar spherical harmonics as
\begin{equation}
F(\theta,\phi,t)
=
\sum_{\ell,m}
a_{\ell m}(t)\,
Y_{\ell m}(\theta,\phi) \, ,
\end{equation}
where $a_{\ell m}$ are the complex multipolar coefficients measured at a given extraction radius.
To estimate the luminosity carried specifically by the polar outflow, we reconstruct the angular distribution of the Poynting flux using the axisymmetric modes up to $\ell=4$,
\begin{equation}
F_{\rm axi}(\theta,t)
=
\sum_{\ell=0}^{4}
a_{\ell 0}(t)\,
Y_{\ell 0}(\theta) \, ,
\end{equation}
which provides an approximate description of the large-scale funnel geometry while avoiding expensive three-dimensional post-processing of the GRMHD data.
The polar luminosity is then defined as the flux integrated over two polar caps with half-opening angle $\theta_0$,
\begin{equation}
L_{\rm polar}(t)=r^2
\int_{\Omega_{\rm polar}}
F_{\rm axi}(\theta,t)\,
d\Omega \, ,
\label{eq:Lpolar}
\end{equation}
where
\begin{equation}
\Omega_{\rm polar}
=
\left\{
\theta < \theta_0
\right\}
\cup
\left\{
\theta > \pi-\theta_0
\right\} \, .
\end{equation}
In practice, we adopt $\theta_0 \sim 17^\circ$, corresponding to the typical opening angle of the low-density, highly magnetized funnel formed after the collapse. When such a magnetically dominated funnel is absent or poorly defined, we instead use $\theta_0 \sim 35^\circ$ to isolate the polar region and exclude contributions from the denser disk material (see Sec.\ref{sec:jet} for more details).


To facilitate comparison with short gamma-ray burst (sGRB) luminosities inferred from observations, we also compute the isotropic-equivalent luminosity,
\begin{equation}
L_{\rm iso}
=
\frac{L_{\rm polar}}
{1-\cos\theta_0} \, ,
\label{eq:Liso}
\end{equation}
where the denominator accounts for the fraction of the sky subtended by the two polar caps with half-opening angle $\theta_0$. This quantity corresponds to the luminosity that would be inferred under the assumption of isotropic emission. The luminosity estimates presented in Sec.~\ref{sec:results} are evaluated at a radius of $r \simeq 443$ km. Both the extraction radius and the definition of $L_{\rm iso}$ follow those adopted by Kiuchi et al.~2024~\cite{Kiuchi-2024-alpha}, enabling a direct comparison with previous estimates of sGRB jet luminosities.

\subsubsection{Other related quantities:}

Spritz outputs density-weighted volume averages of several magnetic-field diagnostics, which we use throughout this work to characterize the magnetic-field evolution. The corresponding local quantities are defined as
\begin{align}
   B_{\rm norm} &= \sqrt{\gamma_{ij} B^i B^j} \, , \\
   B_{\rm tor} &= \sqrt{\left|B^{\phi} B_{\phi}\right|} \, , \\
   B_{\rm pol} &= \sqrt{\left|B_{\rm norm}^2 - B^{\phi} B_{\phi}\right|} \, .
   \label{eq:Bpol_Btor}
\end{align}
%
The reported values correspond to density-weighted volume averages obtained from
\begin{equation}
\langle X \rangle_\rho =
\frac{\int \rho W X \sqrt{\gamma}\, dV}
     {\int \rho W \sqrt{\gamma}\, dV} \, .
\end{equation}

\section{Results}
\label{sec:results}


\subsection{Dynamical Evolution}
\label{sec:dyn_evol}

\begin{figure}
    \centering
    \includegraphics[width=0.999\linewidth]{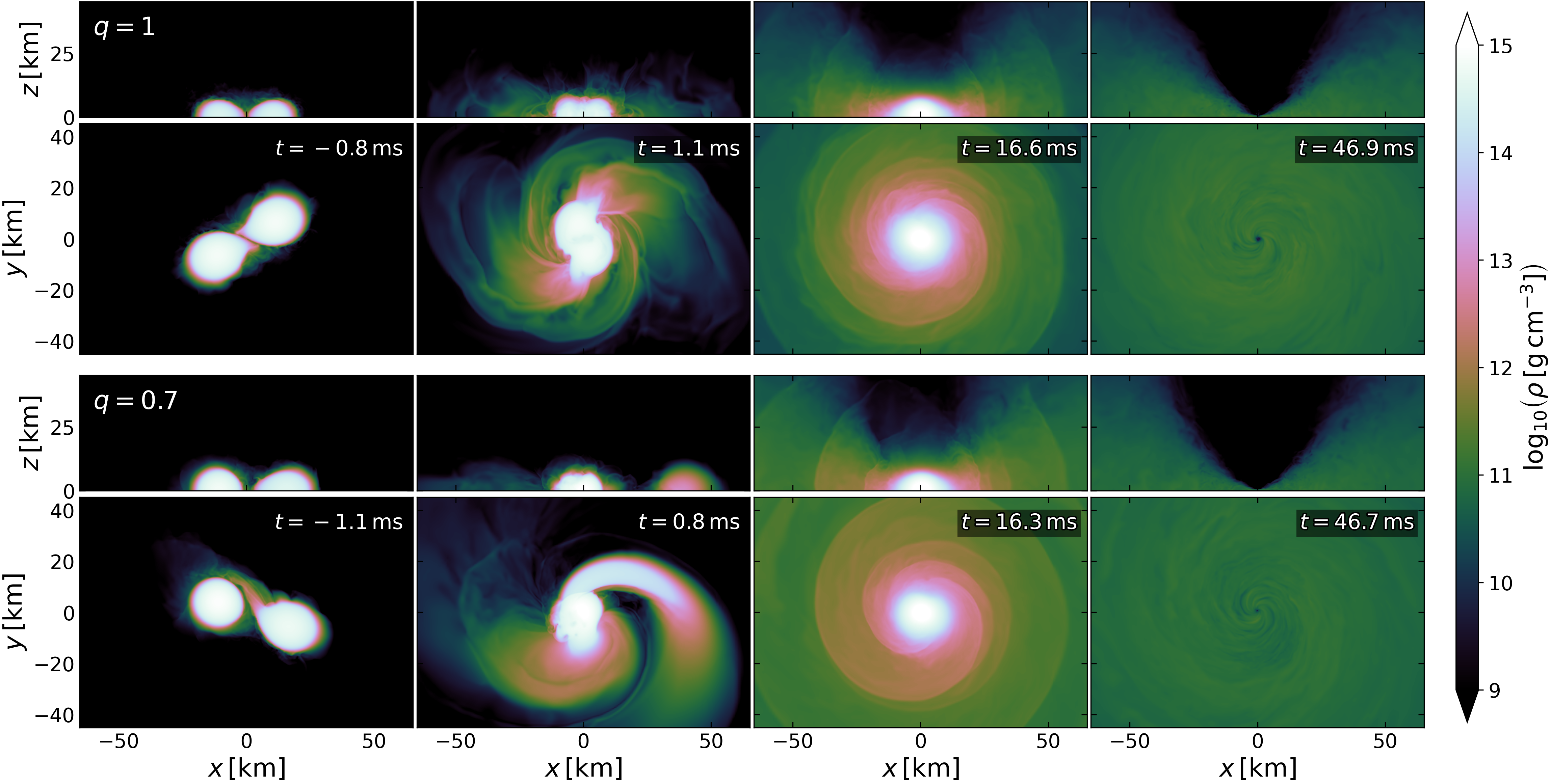}
    \caption{The two-dimensional density profiles from different snapshots for $q=1$ (top) and $q=0.7$ cases (bottom).
    }
    \label{fig:rho-2d}
\end{figure}

\subsubsection{Inspiral, merger and post-merger}

The dynamical evolution of the two binaries is summarized in Fig.~\ref{fig:rho-2d}, which shows density snapshots from the inspiral, merger, HMNS, and BH-disk phases. Despite having identical chirp masses and employing the same EOS, the two systems differ in their mass ratios and consequently exhibit important differences in their merger and post-merger evolutions. During the inspiral, the unequal-mass binary experiences stronger tidal deformation, resulting in a more asymmetric matter distribution near merger and the formation of pronounced tidal tails. The equal-mass and unequal-mass binaries complete approximately 6.2 and 5.8 orbits before merger, respectively.

The merger occurs at $t_{\rm merg}=16.7\mathrm{ms}$ and $15.46\mathrm{ms}$ for the $q=1$ and $q=0.7$ models, respectively. During the merger, strong hydrodynamic shocks develop at the contact interface between the two neutron stars, leading to substantial heating of the remnant and the surrounding material. At the same time, angular-momentum redistribution and tidal torques launch dynamical ejecta into the surrounding environment. Due to its stronger tidal disruption, the unequal-mass binary produces more pronounced tidal tails and a larger amount of equatorially concentrated ejecta, while the equal-mass system exhibits a more symmetric matter distribution. Following the merger, both binaries form a differentially rotating HMNS supported against prompt collapse by thermal pressure and rapid rotation. In the unequal-mass case, the remnant exhibits stronger non-axisymmetric spiral structures and undergoes significant redistribution of angular momentum. The evolution of the minimum lapse function, maximum density, and maximum temperature indicates a gradual contraction of the HMNS over time. 

The thermal evolution of the remnant is primarily driven by shock heating during the merger, the subsequent dissipation of kinetic energy through differential rotation within the HMNS, and compressional heating associated with the gradual contraction of the remnant prior to collapse. The MRI-driven turbulence may provide an additional source of heating during the post-merger evolution.
Shortly after the merger, the maximum temperature rises rapidly as shocks propagate through the remnant and the surrounding material. The temperature continues to increase during the HMNS phase, reaching approximately $150\,\mathrm{MeV}$ and $145\,\mathrm{MeV}$ for the $q=1$ and $q=0.7$ models, respectively, immediately before collapse. In our simulations, the electron fraction is evolved self-consistently while neutrino cooling is neglected. As a result, thermal energy generated by shocks and turbulent motions is not efficiently radiated away, allowing the remnant to remain hotter than would be expected in simulations that include neutrino cooling, see e.g. \cite{Perego:2019adq}. Consequently, the reported temperatures should be interpreted as upper limits to the thermal state of the remnant. 
Furthermore, the electron fraction of each fluid element remains identical to its initial value, due to the absence of neutrino-matter interactions. Given the initial neutron richness of the merging neutron stars, the electron fraction in the remnant and in the ejecta remains significantly low ($Y_e \lesssim 0.1$) due to the missing protonization caused by positron absorption on neutrons and neutrino irradiation, see e.g. \cite{Wanajo:2014wha,Sekiguchi-2015,Radice-2018a}. 

The two remnants differ in their lifetimes. The unequal-mass model collapses to a BH approximately $20.5\,\mathrm{ms}$ after merger, whereas the equal-mass remnant survives for about $31.6\,\mathrm{ms}$. The collapse times obtained in our simulations are consistent with those reported in previous studies using similar binary configurations and the LS220 EOS~\cite{Nedora-2021,Espino-2023}. Monitoring the minimum lapse function, which we use as one of the diagnostics of BH formation, shows the minimum lapse decreases gradually as the HMNS remnant contracts and becomes more compact. Shortly before collapse, the lapse exhibits a rapid decline once it reaches values of $\alpha_{\rm{min}} \sim 0.3$, ultimately dropping to $\alpha_{\rm min}\simeq0.04$ immediately before BH formation. The substantially longer lifetime of the equal-mass remnant allows it to undergo a more extended phase of differential rotation, angular-momentum redistribution, and remnant evolution before collapse.

The substantially longer HMNS lifetime of the equal-mass model provides a longer timescale for differential rotation and magnetic-field amplification, whose consequences are examined in the following sections.

\subsubsection{Black hole-disk system}

Following the collapse of the HMNS, both simulations settle into a BH surrounded by a dense accretion disk (BH-disk system), as illustrated by the final column of Fig.~\ref{fig:rho-2d}. Such BH-disk configurations are widely regarded as promising central engines for short GRBs, providing the conditions required for the extraction of the BH's rotational energy through magnetically driven processes and the launching of relativistic outflows. The post-collapse evolution, therefore, represents a crucial stage linking the merger dynamics to the possible electromagnetic counterparts.

The properties of the newly formed BHs are broadly similar in the two models. For the unequal-mass binary ($q=0.7$), we measure a dimensionless spin parameter of $a_{\rm BH}\simeq0.57$ right after the formation of the apparent horizon, while the equal-mass case yields a slightly larger value of $a_{\rm BH}\simeq0.59$. The corresponding BH masses are $M_{\rm BH}\simeq\,2.42 M_\odot$ and $M_{\rm BH}\simeq\,2.44 M_\odot$, respectively. The similarity of these values reflects the fact that both systems share the same chirp mass and EOS, have a qualitatively similar evolution and start with the same initially non-rotating star configuration, leading to comparable amounts of mass and angular momentum being incorporated into the BH during collapse. 

More obvious differences emerge in the surrounding accretion disks. Measuring the baryonic mass located outside the apparent horizon shortly after BH formation, we estimate upper limits to the disk masses of approximately $M_{\rm disk}\sim 0.3 \,M_\odot$ for the $q=0.7$ model and $M_{\rm disk}\sim 0.18 \,M_\odot$ for the $q=1$ model.
The larger disk mass in the unequal-mass case is expected from the stronger tidal deformation during the merger, which redistributes matter to larger radii and leaves a greater fraction of the remnant outside the horizon after collapse. Consistent with this argument, the density profiles reveal a more extended and massive torus in the $q=0.7$ case, while the equal-mass model forms a comparatively less massive and geometrically more compact disk.

The thermodynamic properties of the post-collapse disks are broadly similar in the two models. Following BH formation, the densest and hottest regions of the remnant are rapidly accreted through the horizon, leaving behind a geometrically thick accretion disk with maximum temperature around $\sim 14 -18 ~\rm MeV$. The maximum density remains above $6 \times 10^{11} \,\rm g/cm^3$ immediately after collapse and gradually decreases by about an order of magnitude during the subsequent evolution. Although no unmagnetized reference simulation is available for direct comparison, previous GRMHD studies have shown that magnetic stresses efficiently transport angular momentum and contribute to expanding the disk and sustaining turbulence and heating in postmerger BH-disk systems~\cite{Nouri:2018}. 

\subsection{Gravitational wave emissions}
\label{sec:GW}

\begin{figure}[t]
  \includegraphics[width=\linewidth]{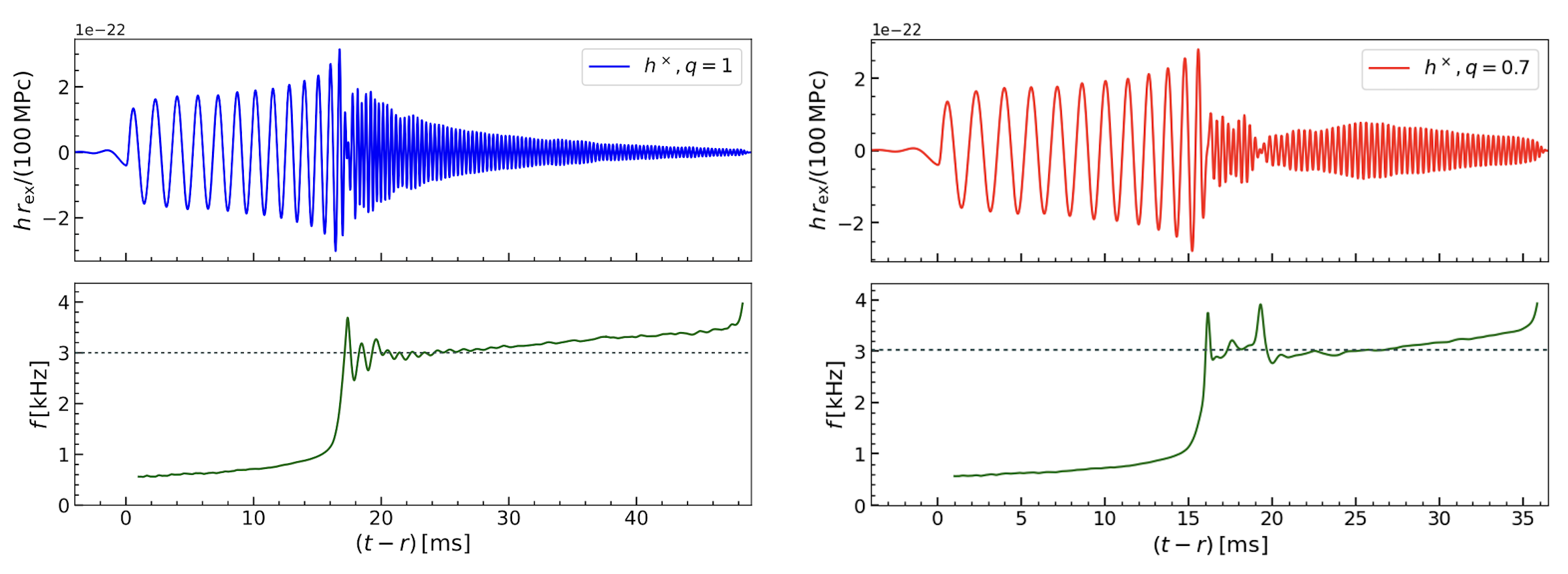}
  \caption[gw-strain]{\justifying (Top) Gravitational wave strain ($h^{x}$), $l=2$, $m=2$ multipole component (Bottom) Instantaneous frequency (phase velocity) of $\Psi_4$. Left panel presents the results from $q=1$ and right panel shows $q=0.7$ cases.}
  \label{fig:gw_strain}
\end{figure}

\begin{figure}
  \includegraphics[width=\linewidth]{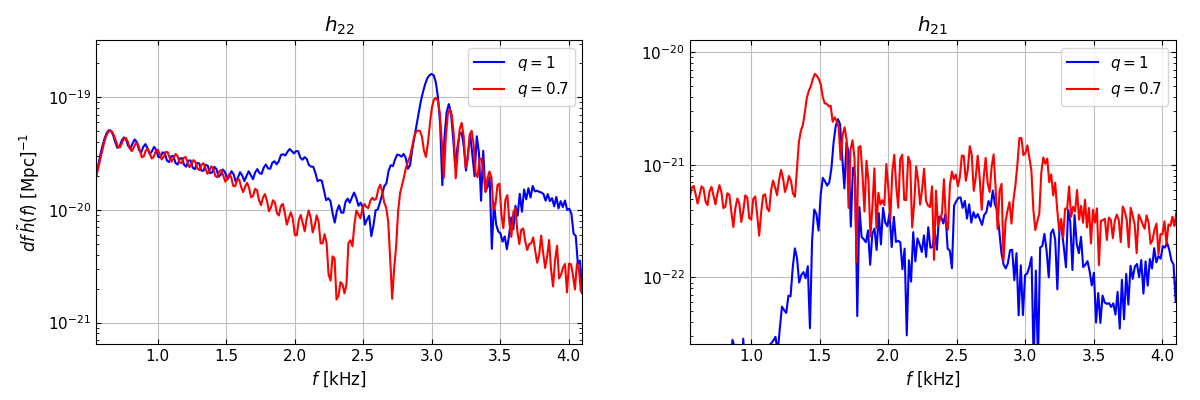}
  \caption[gw-spectra]{\justifying Effective gravitational wave strain spectra for $l=2$, $m=2$ (left) and $l=2$, $m=1$ (right) multipole component, computed from the outermost detector at $d=2510$ km.}
  \label{fig:gw_spectra}
\end{figure}

The GW signal extracted from $\Psi_4$ in the simulations is dominated by the quadrupolar $(\ell,m)=(2,2)$ mode during both the inspiral and post-merger phases, as expected for BNS mergers. Figure~\ref{fig:gw_strain} shows the GW strain together with the instantaneous frequency evolution for the equal-mass ($q=1$) and unequal-mass ($q=0.7$) configurations. In both cases, the inspiral waveform exhibits the characteristic chirping behavior, followed by a sharp amplitude peak at merger and a subsequent oscillatory post-merger signal produced by the differentially rotating HMNS remnant. The postmerger phase is characterized by high-frequency oscillations in the kHz range associated with the fundamental quadrupolar fluid mode of the remnant. The effective strain spectra, shown in Fig.~\ref{fig:gw_spectra}, reveal a dominant peak around $\sim 3\,\mathrm{kHz}$ for both binaries. As illustrated in the figure, changing the mass ratio introduces only a slight shift in the dominant frequency.
Our results are consistent with the findings of Bauswein and Stergioulas (2015)~\cite{Bauswein-Stergioulas:2015}, who showed that unequal-mass binaries exhibit deviations in $f_{\mathrm{peak}}$ of at most a few percent compared to equal-mass systems with similar total masses.

The time evolution of the instantaneous GW frequency further reflects the secular evolution of the HMNS remnant. Following the merger, the dominant frequency gradually increases with time in both simulations. This frequency drift is a consequence of the progressive contraction and densification of the HMNS as angular momentum is redistributed and radiated away through GWs. As the remnant becomes more compact and approaches gravitational collapse, the characteristic oscillation frequencies shift toward higher values. This behavior becomes particularly evident shortly before BH formation, where the frequency evolution steepens significantly. Such a trend is consistent with previous studies of delayed-collapse remnants, including Bernuzzi et al.~(2015) and Breschi et al.~(2024)~\cite{Bernuzzi:2015,Breschi:2024}.

In addition to the dominant $(2,2)$ component, both systems exhibit a clear low-frequency $(2,1)$ mode in the post-merger GW spectrum. This feature is particularly interesting because it is associated with the development of the one-armed spiral, or $m=1$, instability in the HMNS remnant. This instability develops shortly after the merger and is considered one of the effective mechanisms for launching ejecta in long-lived remnants~\cite{Nedora-2021}. 
The corresponding spectral peak appears at frequencies of approximately $1.63\,\mathrm{kHz}$ for the equal-mass binary and $1.46\,\mathrm{kHz}$ for the unequal-mass system (see Fig.~\ref{fig:gw_spectra}). While the $(2,1)$ mode remains subdominant compared to the quadrupolar emission, its amplitude becomes more pronounced in the unequal-mass configuration, consistent with the stronger asymmetry of the remnant and the enhanced excitation of non-axisymmetric structures, in agreement with previous studies, such as Lehner et al. (2016)~\cite{Lehner:2016}. The presence of this mode indicates the persistence of a dynamically active HMNS over several milliseconds after merger and may provide an additional observational signature in the post-merger GW spectrum. In the equal-mass case, the instability can still develop through mechanisms that break the initial symmetry of the system. One such mechanism is the KHI, arising in the shear layer formed at the contact interface during merger~\cite{Lehner:2016,Anderson:2008}.

\subsection{Magnetic field evolution}
\label{sec:B_evol}

\subsubsection{Magnetic field amplification}

The amplification of magnetic fields during and after merger plays a central role in the long-term evolution of the remnant and the possible formation of relativistic jets and outflows. Although the binaries are initialized with strong purely poloidal magnetic fields $B_{\rm{max}}\simeq10^{16} \mathrm{G}$, the merger and post-merger processes trigger amplification mechanisms, including KHI at the shear interface between the merging stars, magnetic winding driven by differential rotation, and MRI in the differentially rotating remnant and surrounding disk. In this section, we examine the evolution of the magnetic field components, electromagnetic energy, and rotational structure of the remnant.

\begin{figure}
    \centering
    \includegraphics[width=0.999\linewidth]{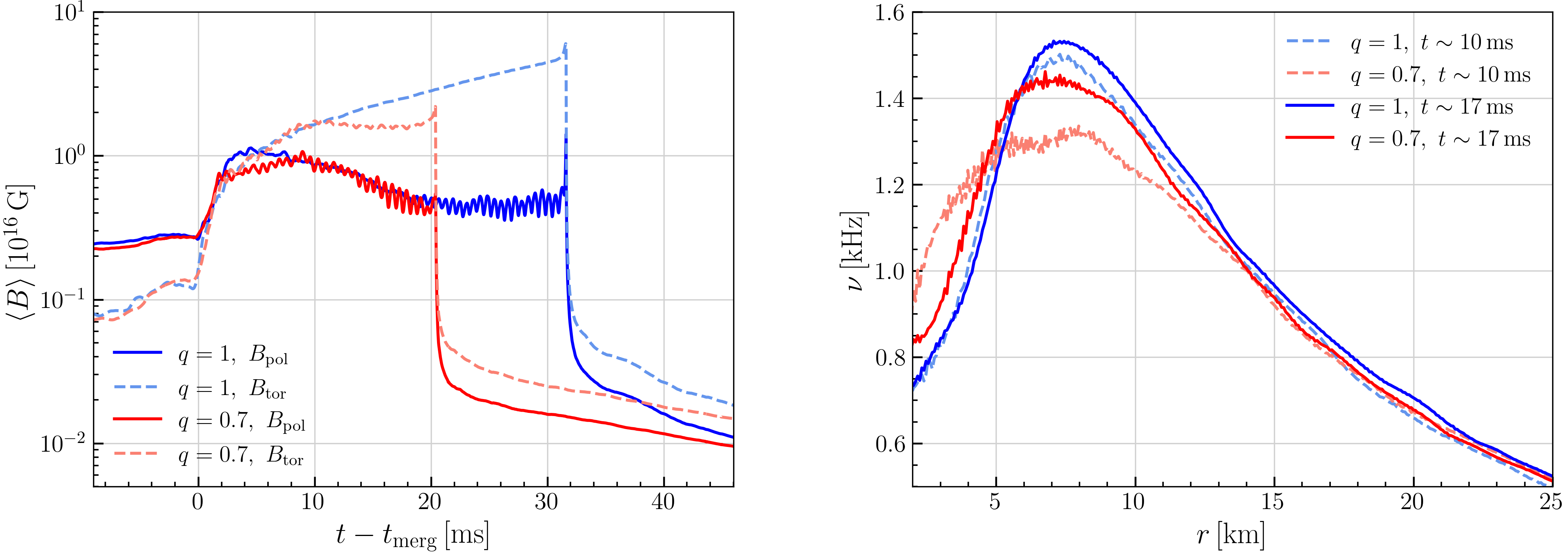}
    \caption{\justifying The evolution of the toroidal and poloidal components of the magnetic field (left), and the rotation profiles at different time snapshots for the HMNS and the envelope (time labels show time after the merger).}
    \label{fig:B_pol-B-tor}
\end{figure}

The left panel of Fig.~\ref{fig:B_pol-B-tor} shows the evolution of the density-weighted volume-averaged poloidal and toroidal magnetic field components (Eq.~\ref{eq:Bpol_Btor}). Prior to the merger, the magnetic field is predominantly poloidal with a characteristic strength of $<B_{\rm pol}>\simeq 2 \times 10^{15}\,\mathrm{G}$, while the toroidal component remains comparatively weak. Shortly after the merger, both components are rapidly amplified, reaching strengths of order $10^{16}\,\mathrm{G}$. The amplification is initially driven by the strong shear layer formed at contact between the two neutron stars, i.e.~the KHI. Following the merger, the toroidal component continues to grow and eventually exceeds the poloidal component by several factors. This behavior is consistent with magnetic winding in the differentially rotating HMNS. In contrast, the poloidal component reaches a maximum shortly after merger and subsequently decreases. Small oscillations are visible in the poloidal-field diagnostic during the late HMNS phase; however, these oscillations become significantly weaker at higher resolution and are therefore not interpreted as a physical effect (see Appendix~\ref{sec:resolution_study} for the resolution study); instead, they are likely of numerical origin.

The evolution of the toroidal field can be understood in the context of the rotational structure of the remnant. The right panel of Fig.~\ref{fig:B_pol-B-tor} shows the angular-velocity profiles at $t=10\,$ms and $t=17$\,ms after the merger, during the HMNS phase. In both models, the remnant exhibits strong differential rotation, providing favorable conditions for magnetic winding and the conversion of poloidal magnetic field into toroidal field. As the remnant evolves, the peak angular velocity gradually increases as a consequence of the progressive contraction of the HMNS. This behavior contrasts with the long-lived remnant studied by Ciolfi et al.~(2019)~\cite{Ciolfi-2019}, where the angular velocity decreases with time during the post-merger evolution. 
Although the rotational profiles of the two models are broadly similar, the equal-mass remnant survives approximately $11~\mathrm{ms}$ longer before collapse. This extended HMNS phase allows magnetic winding amplification to continue operating for a longer period, resulting in larger toroidal field strengths compared to the unequal-mass model. In contrast, the toroidal component in the $q=0.7$ case begins to level off shortly before collapse, suggesting that the magnetic field amplification approaches saturation as angular momentum redistribution and magnetic backreaction gradually reduce the efficiency of further winding. 

\begin{figure}[t]
\centering
  \includegraphics[width=0.8\linewidth]{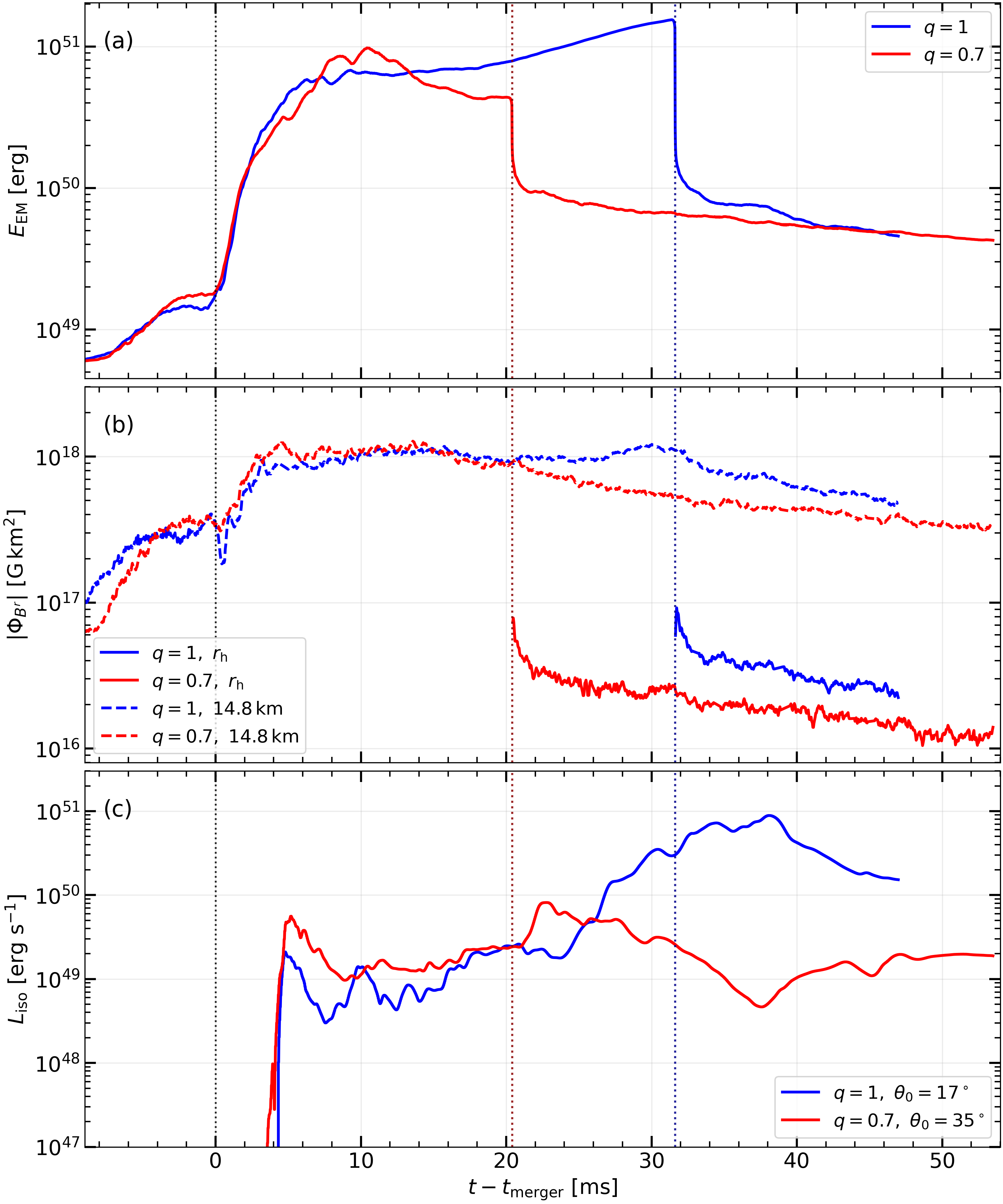}
  \caption[E-evolution]{\justifying The evolution of the total electromagnetic energy (top), radial magnetic flux (middle), and the isotropic equivalent Poynting-flux luminosity (bottom).}
  \label{fig:Eem-flux-L}
\end{figure}

The growth of the magnetic field is also reflected in the evolution of the total electromagnetic energy (Eq.~\ref{eq:E_em}) shown in Fig.~\ref{fig:Eem-flux-L}(a). In both simulations, the electromagnetic energy increases by nearly two orders of magnitude following merger, rising from $\sim\!10^{49}\,\mathrm{erg}$ to values approaching $10^{51}\,\mathrm{erg}$. This rapid increase is driven initially by KHI and subsequently by magnetic winding and MRI-induced turbulence within the differentially rotating remnant. Consistent with the evolution of the toroidal magnetic field, the equal-mass model continues to accumulate magnetic energy throughout its longer HMNS lifetime, ultimately reaching larger electromagnetic energies prior to collapse. In contrast, the electromagnetic energy in the unequal-mass model reaches its maximum well before collapse and subsequently decreases during the remaining HMNS phase. This behavior suggests that magnetic field amplification becomes progressively less efficient and is eventually overtaken by magnetic energy dissipation and/or redistribution within the remnant. Such an interpretation is consistent with the saturation of the toroidal component and the decline of the poloidal component observed in Fig.~\ref{fig:B_pol-B-tor}.

Following BH formation, the electromagnetic energy exhibits a sharp decrease in both models as a substantial fraction of the strongly magnetized HMNS is accreted through the horizon. Subsequently, the magnetic energy evolution becomes comparatively flat, and no further significant amplification is observed.
Although the available reservoir of differential rotational energy is substantially reduced after collapse, MRI-driven turbulence may still operate within the accretion disk around the formed BH. Despite having different HMNS histories, both models converge toward similar electromagnetic energies of $\approx 4 \times10^{49}\,\mathrm{erg}$.

\subsubsection{Magnetic-flux accumulation}

Figure~\ref{fig:Eem-flux-L}(b) shows the evolution of the radial magnetic flux measured both on the apparent horizon (solid lines) and on a spherical surface located at $r\simeq14.8~\mathrm{km}$ (dashed lines). While the volume-averaged electromagnetic energy quantifies the overall magnetic energy content of the remnant, the mean-radial magnetic flux measured according to Eq.~(\ref{eq:Phi_Br}) provides information about the large-scale poloidal field around the central compact remnant. In the context of jet formation, the accumulation of coherent poloidal magnetic flux around the central object is particularly important, as it determines the efficiency with which rotational energy can be extracted and converted into electromagnetic luminosities.

Before collapse, both models exhibit a significant increase in radial magnetic flux, reflecting the amplification of the magnetic field during merger and the HMNS phase. The flux measured at $r\simeq14.8\,\mathrm{km}$ reaches values of order $10^{18}~\mathrm{G\,km^2}$ and remains relatively stable until BH formation. Consistent with the behavior of the electromagnetic energy, the equal-mass model maintains slightly larger magnetic fluxes towards the end of the HMNS phase. Following collapse, the magnetic flux decreases in both cases as highly-magnetized material is accreted by the BH. Looking at the magnetic flux measured on the horizon, after collapse, we observe values about 1.5 orders-of-magnitude smaller and a similar decreasing trend.

\subsection{Incipient jet}
\label{sec:jet}

A central question in the post-merger evolution of BNS mergers is whether the remnant develops conditions favorable for launching a relativistic jet. In magnetically driven models of sGRBs, the formation of a relativistic jet generally requires a rapidly rotating compact object threaded by a large-scale poloidal magnetic field, together with a sufficiently baryon-poor polar region through which the outflow can propagate. The previous sections demonstrated that the equal-mass remnant accumulates larger magnetic energies and magnetic fluxes during its longer HMNS lifetime. We now investigate whether these differences translate into distinct funnel structures and/or collimated polar outflow in the post-collapse phase.

\subsubsection{Highly magnetized funnel formation} 

\begin{figure}[t]
    \centering
    \includegraphics[width=0.95\linewidth]{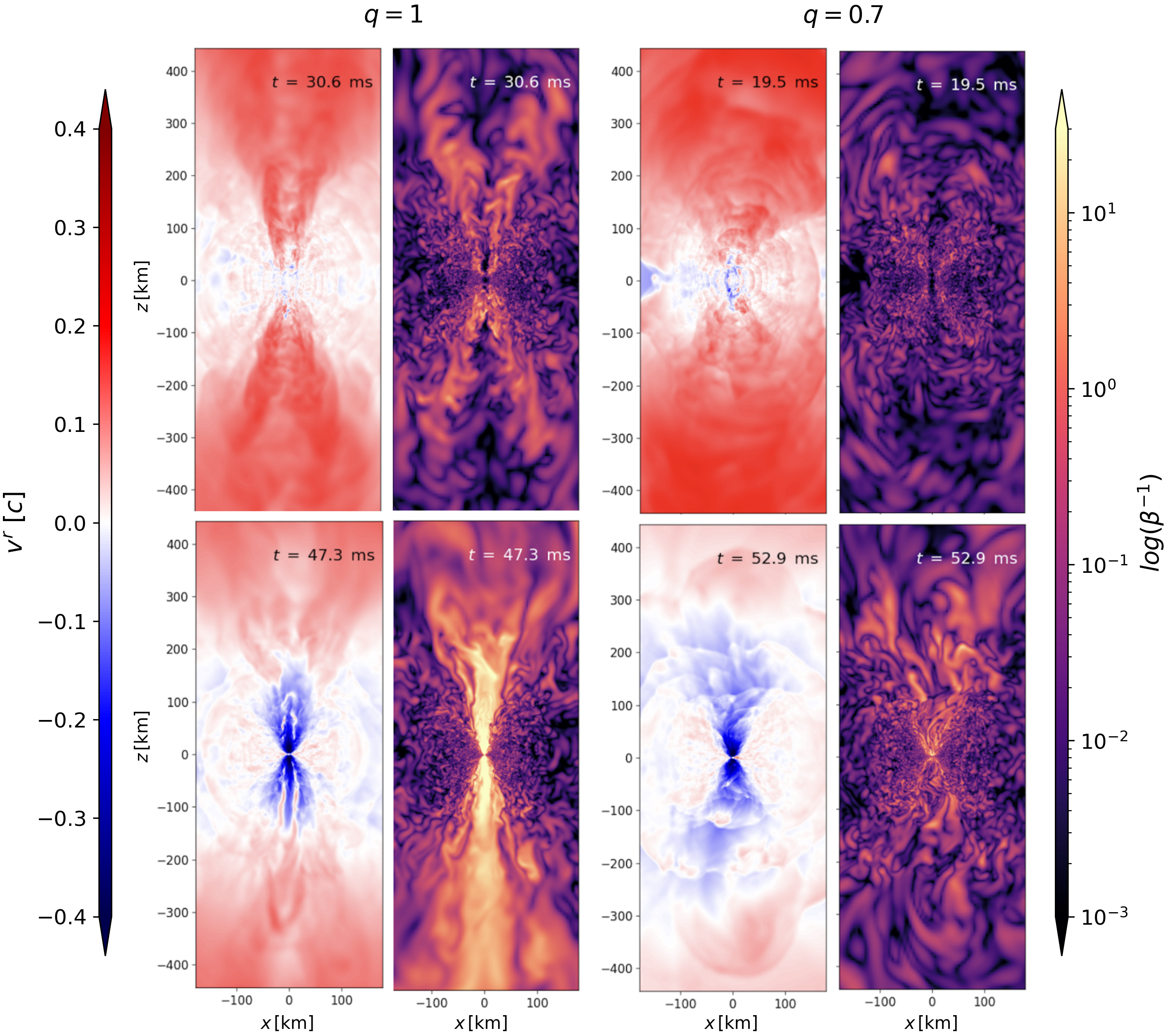}
    \caption{\justifying 2D profiles of radial velocity and $\beta^{-1}$ at $\sim 1$ ms before the collapse (top) and at the end of the simulations (bottom). 
    }
    \label{fig:2d-vr-beta}
\end{figure}

Figure~\ref{fig:2d-vr-beta} shows the two-dimensional distributions of the radial velocity and the inverse plasma beta parameter, defined as the ratio of the magnetic pressure to the gas pressure, $\beta^{-1}= P_B/P_g$.
Regions with $\beta^{-1}>1$ are magnetically dominated and are generally regarded as favorable sites for the launching of magnetically driven jets. In particular, the formation of a coherent magnetically dominated funnel along the rotation axis is considered one of the essential ingredients for the development of a Blandford-Znajek type jet. The snapshots of the 2D profiles are taken at $\sim 1~\mathrm{ms}$ before the collapse and at the end of simulations, which are $\sim 15~\mathrm{ms}$ and $\sim 30~\mathrm{ms}$ after the collapse for the equal-mass and unequal-mass models, respectively. 

Before collapse, both models exhibit magnetized polar outflows emerging from the HMNS. In the equal-mass case, regions with $\beta^{-1} \gtrsim 1$ begin to develop above the remnant poles, indicating that magnetic stresses have become locally comparable to or larger than the gas pressure. However, the polar environment remains significantly contaminated by baryonic material and the magnetically dominated regions are confined to relatively small scales. The unequal-mass model displays a qualitatively similar behavior, although the magnetized polar structures appear weaker and less organized. At this stage, neither remnant exhibits a well-defined funnel extending to large distances from the central object.

The most significant differences appear after BH formation. In the equal-mass model, a coherent magnetically dominated funnel develops along the rotation axis and expands to distances of a few hundred kilometers above the BH. The funnel is characterized by extended regions with $\beta^{-1}>1$, indicating that magnetic pressure dominates over gas pressure throughout a large fraction of the polar region. Simultaneously, the radial velocity maps reveal outward-moving material aligned with the rotation axis, suggesting the presence of a collimated magnetized outflow. 
In contrast, the unequal-mass model develops a significantly weaker polar structure. Although localized magnetically dominated regions are present, they remain less extended and more fragmented than in the equal-mass case. The corresponding radial velocity distribution appears more turbulent and less collimated, indicating that the outflow remains dominated by a combination of magnetized winds and turbulent motions.

One possible interpretation of the weaker polar structure in the unequal-mass model is that the remnant simply requires a longer post-collapse evolution to develop a magnetically dominated funnel. To investigate this possibility, the $q=0.7$ simulation was evolved for approximately $\sim 30~\mathrm{ms}$ after BH formation, roughly twice as long as the post-collapse evolution followed in the equal-mass case. Despite this extended evolution, no coherent large-scale funnel comparable to that observed in the $q=1$ model appears. This suggests that the development of a well-defined funnel is delayed in the unequal-mass model, likely reflecting differences in the magnetic-field evolution established during the HMNS phase, although longer and higher-resolution simulations would be required to determine whether a comparable funnel eventually develops.

The weaker funnel formation in the unequal-mass model may also be related to the intrinsically more asymmetric post-merger configuration. As discussed in Sec.~\ref{sec:dyn_evol}, the $q=0.7$ merger produces stronger tidal deformation and a more extended disk. The resulting non-axisymmetric structure can continuously redistribute baryonic material into the polar regions, making it more difficult to establish and maintain a clean magnetically dominated funnel. This interpretation is further supported by the time evolution of the density distribution, which reveals episodic injections of matter into the polar region throughout the post-collapse evolution of the unequal-mass model. Such episodic baryon loading delays the establishment of a persistent low-density funnel and may contribute to the suppression of jet formation.

\subsubsection{Polar baryon pollution} 

\begin{figure}
    \centering
    \includegraphics[width=0.8\linewidth]{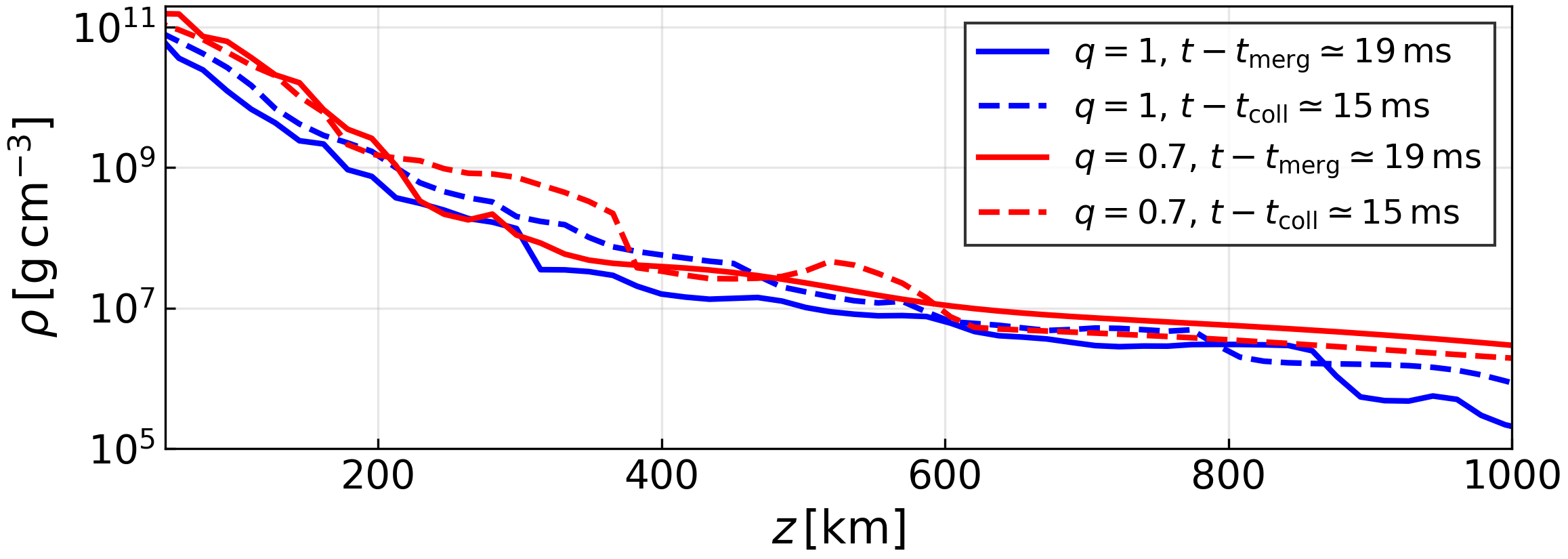}
    \caption{\justifying The density profile as a function of distance from the remnant along the polar direction at two different snapshots: $t \sim 19$ms after the merger (solid line), and $t \sim 15$ms after the collapse (dashed line).
    }
    \label{fig:rho-vs-z}
\end{figure}

To quantify the degree of baryon pollution in the polar regions, Fig.~\ref{fig:rho-vs-z} shows the density profile along the $z$ axis for both simulations at two representative evolutionary stages: approximately $19\,\mathrm{ms}$ after merger and $15\,\mathrm{ms}$ after BH formation.
We observe that at all evolutionary stages the equal-mass model maintains lower densities along the polar direction than the unequal-mass case, consistent with the cleaner funnel structure observed in Fig.~\ref{fig:2d-vr-beta}. The higher densities in the $q=0.7$ model likely originate from its more asymmetric matter distribution, and the episodic redistribution of material into the polar region observed during the post-collapse evolution. Nevertheless, the difference between the two models generally remains below one order of magnitude. 
An interesting observation of both models is that the polar density remains relatively high after BH formation. This suggests that the collapse itself does not immediately produce a clean funnel, despite the appearance of magnetically dominated regions in Fig.~\ref{fig:2d-vr-beta}.
Therefore, neither remnant develops a considerably evacuated polar funnel. Significant baryonic material remains present along the axis even after BH formation, indicating that both systems remain largely baryon loaded.

Overall, the lower polar densities in the equal-mass model provide a more favorable environment for a magnetically driven outflow. However, since the difference in baryon loading remains modest, baryon pollution alone is unlikely to account for the substantially different funnel structures.

\subsubsection{Electromagnetic luminosity} 
\label{EM-Lum}

To quantify the electromagnetic power associated with the polar region, we consider the isotropic-equivalent luminosity $L_{\rm iso}$ defined in Eq.~(\ref{eq:Liso}). The computation requires specifying a characteristic opening angle $\theta_0$ over which the polar luminosity is integrated. Inspection of the funnel morphology shown in Fig.~\ref{fig:2d-vr-beta} indicates that the two models develop significantly different polar structures. For the equal-mass remnant, the magnetically dominated region remains relatively well collimated, allowing us to adopt $\theta_0=17^\circ$. In contrast, the weaker and more turbulent funnel in the unequal-mass model extends over a substantially broader angular region. For this reason, we adopt a larger opening angle of $\theta_0=35^\circ$ when evaluating its isotropic-equivalent luminosity. Smaller values of $\theta_0$ in the unequal-mass case lead to significant cancellations in the spherical-harmonic reconstruction of the polar luminosity and do not adequately capture the full extent of the outflow. The chosen values, therefore, provide a more representative estimate of the electromagnetic power associated with the polar regions in each model.

Figure~\ref{fig:Eem-flux-L}(c) shows the evolution of $L_{\mathrm{iso}}$ extracted at $r=443\,\mathrm{km}$ for both models. The equal-mass model ultimately reaches isotropic-equivalent luminosities approaching $10^{51}~\mathrm{erg/s}$ whereas the unequal-mass model remains significantly dimmer. This contrast mirrors the differences observed in the magnetic energy evolution and funnel morphology observed in previous figures. The stronger and more coherent magnetic field structure established during the longer HMNS phase of the equal-mass remnant appears to translate directly into a more powerful electromagnetic luminosity. 

Both models demonstrate a decline in the measured Poynting-flux luminosity approximately $5-10\,\mathrm{ms}$ after BH formation. A similar behavior has been reported in simulations of short-lived merger remnants by Kalinani et al. (2026)~\cite{Kalinani-2026}, who found that the electromagnetic luminosity initially decreases after collapse, before increasing again at later times. In their simulations, the late-time luminosity eventually approaches a value consistent with that predicted by the Blandford-Znajek mechanism. 
Although our simulations do not extend sufficiently far into the post-collapse phase to determine whether a similar transition occurs, the equal-mass model exhibits several features consistent with the conditions required for activation of this mechanism. These include the presence of a rotating BH, large-scale poloidal magnetic flux, and a magnetically dominated polar region. In addition, the equal-mass model forms a slightly more rapidly rotating BH ($a_{\rm BH}\simeq0.59$) than the unequal-mass model ($a_{\rm BH}\simeq0.57$). Since the Blandford-Znajek luminosity scales as $L_{\rm BZ}\propto a_{\rm BH}^{2}\Phi_{B}^{2}$, both the enhanced magnetic flux accumulation observed in Fig.~\ref{fig:Eem-flux-L}(b) and the larger BH spin act in the same direction, favoring the jet launching conditions in the equal-mass remnant.

\subsection{Ejecta}
\label{sec:ejecta}

\subsubsection{Total mass ejection and mass-loss rates} 

\begin{figure}
  \includegraphics[width=\linewidth]{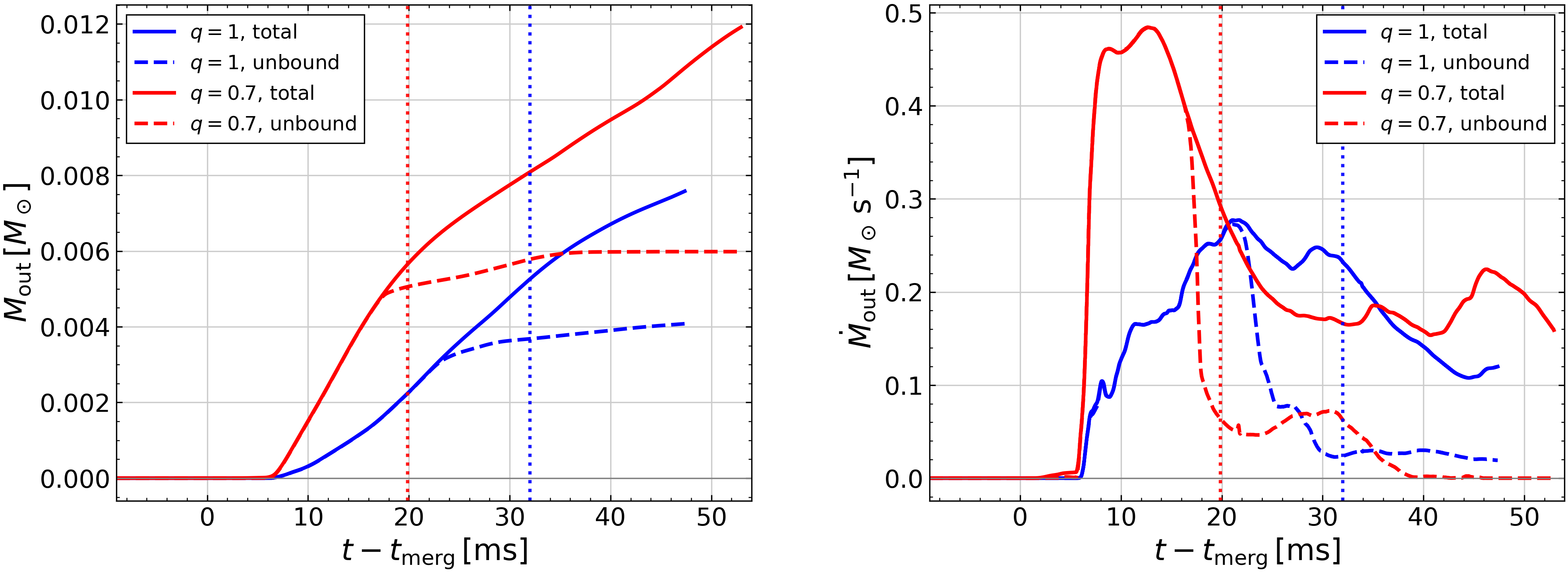}
  \caption[ejecta]{The total outflow and unbound outflow (geodesic criterion) and their rates measured at $r=738$km.}
  \label{fig:1D-ejecta}
\end{figure}

Figure~\ref{fig:1D-ejecta} presents the cumulative total outflow mass and the corresponding mass-loss rates measured at $r\simeq738 \,\mathrm{km}$, together with the cumulative geodesic-unbound $u_t<-1$ ejecta mass and its mass-loss rate. 
The cumulative outflow mass begins to increase shortly after merger in both simulations and continues to grow throughout the HMNS phase and after BH formation. The unequal-mass binary consistently ejects more material than the equal-mass model. By the end of the simulations, the total outflow reaches approximately $1.2\times10^{-2}M_\odot$ for the unequal-mass model and $7.6\times10^{-3}M_\odot$ for the equal-mass binary. The corresponding geodesic-unbound masses are approximately $6\times10^{-3}M_\odot$ and $4\times10^{-3}M_\odot$, respectively. 

The mass-loss rates exhibit an initial peak shortly after merger, associated with the dynamical ejecta generated by tidal torques and shock heating. Then it is followed by a gradual decline during the subsequent evolution. The modest increase in the outflow rate during the HMNS phase, most visible in the $q=0.7$ model, may be associated with non-axisymmetric remnant dynamics, in particular the enhanced $m=1$ spiral instability. Such a mode can transport angular momentum and launch additional material from the remnant. This interpretation is consistent with the stronger $(2,1)$ GW component in the unequal-mass model observed in Fig.~\ref{fig:gw_spectra}. After BH formation, the mass-loss rate decreases in both models, although a non-negligible outflow persists throughout the remainder of the simulations, indicating that mass ejection continues after collapse. 

An interesting feature of Fig.~\ref{fig:1D-ejecta} is the difference between the total outflow and the geodesic-unbound component. In both simulations, roughly one-half of the material crossing the extraction sphere already satisfies the geodesic criterion. For large enough radii, the outflow mass provides a very good estimate of the ejecta evaluated using the Bernoulli criterion ($-hu_t>1$, where $h$ is the specific relativistic enthalpy). Since both the usage of the Bernoulli criterion and the lack of neutrino cooling  generally yield larger ejecta masses than the more restrictive geodesic criterion~\cite{Foucart:2021}, the physically unbound ejecta are expected to lie between the two measurements presented here.

The continued growth of both the total outflow and the geodesic-unbound component during the first $\sim 10-15$ms postmerger evolution suggests that the ejecta are progressively accelerated after their initial launch. The dynamical ejecta is generated through tidal torques and shock heating during merger, whereas the subsequent evolution is likely influenced by additional acceleration mechanisms operating in the remnant. These include angular momentum transport associated with spiral waves in the non-axisymmetric HMNS and magnetically driven turbulence powered by MRI. Together, these processes can continue to transfer energy and angular momentum to the outflowing material, increasing the fraction of matter that eventually becomes gravitationally unbound.

\subsubsection{Ejecta and jet environment}

\begin{figure}
    \centering
    \includegraphics[width=0.85\linewidth]{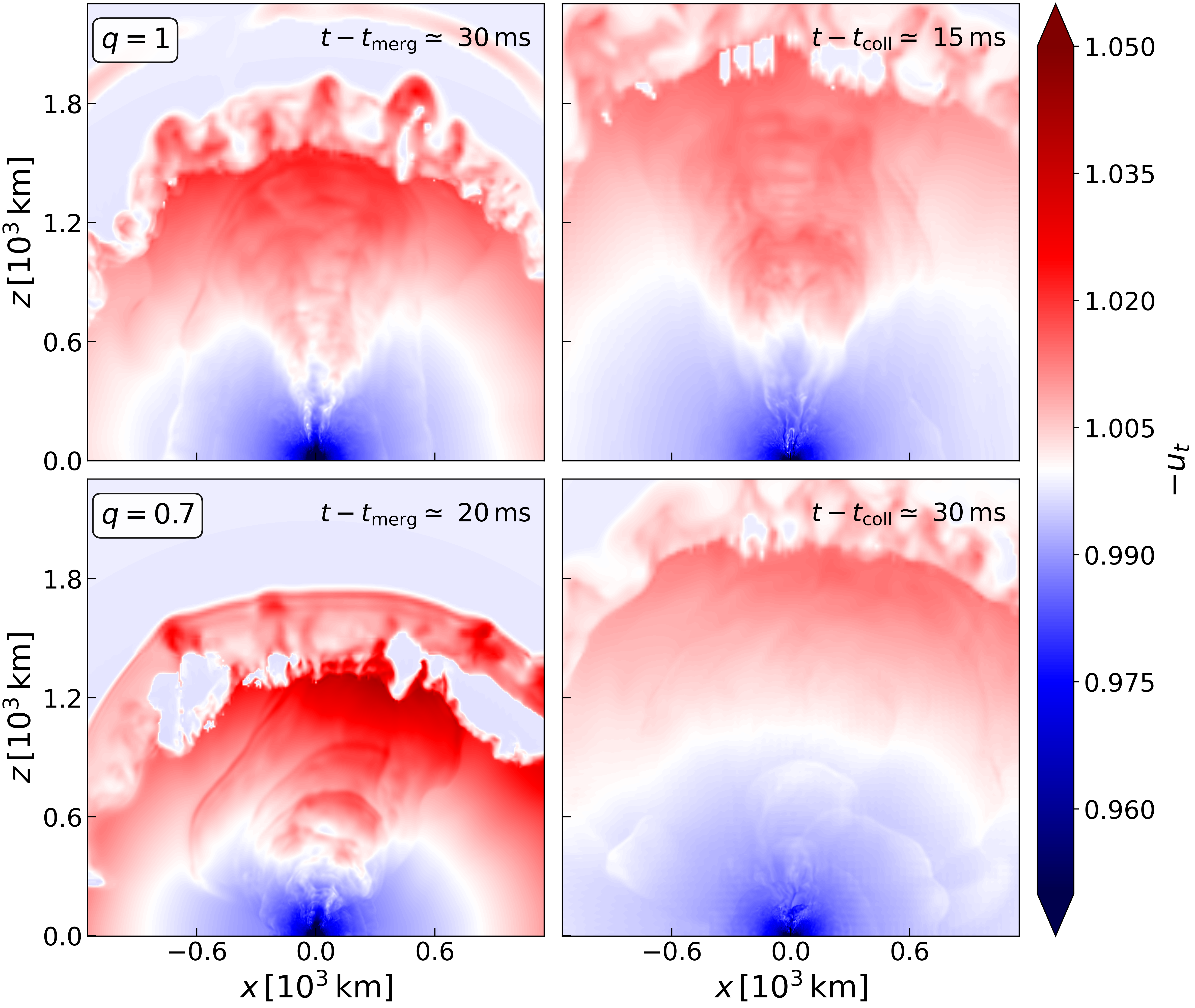}
    \caption{\justifying The unbound mass (red) as identified by geodesic criterion. The snapshots are taken from $\sim 1$ms before the collapse (left) and at the end of the simulations (right).}
    \label{fig:2D-ejecta}
\end{figure}

Figure~\ref{fig:2D-ejecta} presents the two-dimensional distribution along the $y=0$ plane of the unbound material identified by the geodesic criterion at the same evolutionary stages considered in Fig.~\ref{fig:2d-vr-beta}, shortly before BH formation and at the end of the simulations. The ejecta morphology shown here provides complementary information on the environment into which any future relativistic outflow would propagate. We emphasize that our simulations do not follow the long-term evolution of a relativistic jet; instead, they characterize the conditions established by the remnant prior to any possible jet breakout. 

The results in Fig.~\ref{fig:2D-ejecta} should be interpreted together with the radial-velocity maps shown in Fig.~\ref{fig:2d-vr-beta}. In the equal-mass model, the magnetically dominated polar region is less baryon-loaded, favoring the outward propagation of material expelled from the central regions. The outward radial velocities in this region reach values of the order $v_r\sim0.2$c, and the corresponding unbound material is already visible close to the central remnant. This suggests that magnetic stresses within the funnel and along its boundaries contribute to accelerating part of the baryon-loaded polar outflow to geodesically unbound energies. In the unequal-mass model, by contrast, the outflow is broader, more asymmetric, and less coherently aligned with the polar direction, so the unbound material appears less connected to a well-defined axial acceleration channel.

Although our simulations do not extend sufficiently far to follow the subsequent propagation of a relativistic jet, the ejecta morphology provides useful insight into the environment through which such a jet would evolve. In the equal-mass model, the combination of a magnetically accelerated polar outflow and a magnetically dominated funnel suggests that a future jet would encounter less resistance as it propagates through the surrounding ejecta, potentially favoring successful breakout. In contrast, the broader ejecta distribution, the lack of a strong magnetically dominated funnel structure, and sustained baryon loading in the unequal-mass model would likely enhance the interaction between the outflow and the surrounding material, increasing the amount of energy deposited into the ejecta before any possible breakout. Whether this interaction ultimately leads to a successful breakout, a choked outflow, or the formation of a cocoon requires longer simulations, which lie beyond the scope of the present work.

As the ejecta propagate to increasingly larger distances, the influence of the constant-density artificial atmosphere becomes more important. Therefore, although the present simulations reliably capture the early mass ejection and the near-remnant outflow structure, they are not intended to provide quantitatively accurate predictions for the long-distance evolution of the ejecta or its subsequent interaction with a relativistic jet. Addressing this regime will require a more realistic atmosphere prescription, such as the radially decreasing power-law atmosphere adopted by Kalinani et al. (2026)~\cite{Kalinani-2026}, which we plan to implement in future work.

\subsection{Comparison with previous studies}

\subsubsection{Jets and collimated outflows}

Our results can be directly compared with previous GRMHD simulations of delayed-collapse BNS mergers employing initially confined poloidal magnetic fields.
Similar to these studies, we find that the HMNS phase is characterized by significant magnetic field amplification, primarily driven by magnetic winding before collapse. In agreement with, e.g., Kawamura et al.~(2016)~\cite{Kawamura-2016}, the longer-lived remnant develops stronger magnetic fields before BH formation. However, unlike the simulations of Kawamura et al.~\cite{Kawamura-2016} or, e.g., Ciolfi et al.~(2017)~\cite{Ciolfi-2017}, where the post-collapse polar region remained either insufficiently magnetized or failed to develop clear signatures of an incipient jet within the simulated times, our equal-mass model rapidly forms a magnetically dominated polar funnel within only $\sim 15$ ms after BH formation. 
We note, however, that our simulations adopt an initial magnetic field approximately one order of magnitude stronger than those of Ciolfi et al.~(2017) and about six times stronger than that of Kawamura et al.~(2016), which likely contributes to the more rapid development of the post-collapse magnetic funnel.
In addition, these results are broadly consistent with the conclusions of Ruiz et al.~\cite{Ruiz-2021}, who demonstrated that delayed-collapse BH-disk systems can develop magnetically powered incipient jets even when the magnetic field is initially confined inside the neutron stars.

Our equal-mass model also shares several characteristics with the long-lived magnetar simulations of Kiuchi et al.~(2024)~\cite{Kiuchi-2024-alpha}, including stronger magnetic-flux accumulation and enhanced Poynting-flux luminosities. Nevertheless, there are still important differences. Their simulations follow a long-lived HMNS that develops an MRI-driven large-scale magnetic field, likely through an $\alpha\Omega$ dynamo, ultimately producing a Poynting-flux-dominated relativistic outflow. Our HMNS collapses after only $\sim 31$ ms, preventing the development of such a long-lived magnetar phase. Moreover, our resolution is insufficient to fully resolve the MRI and the possible dynamo mechanism. We therefore interpret the enhanced magnetic flux in our equal-mass model primarily as the result of the survival and winding of the initial large-scale poloidal field rather than as evidence for a developed dynamo.

An interesting comparison can also be made with the long-lived magnetar simulations of Ciolfi et al.~(2019) and Ciolfi (2020)~\cite{Ciolfi-2019,Ciolfi-2020}. Those studies demonstrated that magnetic energy can accumulate to $\sim 10^{51}$\,erg and eventually produce a magnetically driven, collimated outflow. However, this outflow develops only after $\sim 100-170$\,ms of postmerger evolution and is driven primarily by magnetic pressure gradients generated by differential rotation in the long-lived remnant. By contrast, our equal-mass delayed-collapse model develops a magnetically driven polar outflow with characteristic velocities of $\sim 0.2\,c$ within only $\sim 30$\,ms after merger, coincident with the formation of a magnetically dominated funnel shortly after BH formation. Similar to the findings of Ciolfi et al., we find that the funnel structure provides conditions favorable for accelerating baryonic material and launching gravitationally unbound polar outflows at relatively small distances from the remnant. Following BH formation, the emergence of a magnetically dominated funnel suggests that the post-collapse configuration is consistent with the early stages of a Blandford-Znajek-powered jet in a BH-disk system. Taken together, our results indicate that the equal-mass model may undergo a transition from a magnetar-launching mechanism to the onset of a Blandford-Znajek launching mechanism after collapse.

A particularly relevant comparison can be made with the recent delayed-collapse simulations of Kalinani et al.~(2026)~\cite{Kalinani-2026}, who investigated the effect of the HMNS lifetime by varying the collapse time in otherwise identical equal-mass binaries. Similar to our simulations, they find that BH formation is followed by the development of a magnetically dominated polar funnel, whose magnetization decreases with increasing HMNS lifetime owing to enhanced baryon pollution. In our models, however, the difference between the equal- and unequal-mass binaries appears to be governed primarily by the asymmetric merger dynamics associated with the binary mass ratio rather than by the remnant lifetime alone. The unequal-mass merger produces a more asymmetric remnant together with more extended tidal ejecta and a denser polar environment, leading to a weaker magnetically dominated funnel despite its shorter HMNS lifetime. In contrast, although BH formation in our equal-mass model does not completely evacuate the polar region, it develops a slightly cleaner and more strongly magnetized funnel than the unequal-mass case. Our comparison therefore suggests that the HMNS lifetime alone is not sufficient to determine the conditions for jet launching. Instead, the formation of a magnetically dominated funnel is governed by the competition between pre-collapse magnetic field amplification and baryon pollution, both of which are influenced by several effective parameters, including the binary mass ratio, EOS, neutrino physics, initial magnetic field configuration, and the remnant's lifetime.

\subsubsection{Ejecta features}

Our ejecta properties are broadly consistent with previous numerical studies showing that unequal-mass mergers produce more massive ejecta and accretion disks than equal-mass binaries. Hydrodynamic simulations by Hotokezaka et al.~(2013)~\cite{Hotokezaka-2013} first demonstrated that increasing the mass asymmetry enhances the dynamical ejecta through stronger tidal disruption of the less massive neutron star. Similar trends were subsequently reported by, e.g., Lehner et al.~(2016)~\cite{Lehner:2016} and Ciolfi et al.~(2017)~\cite{Ciolfi-2017}. We recover the same qualitative behavior: the $q=0.7$ model ejects approximately 50\% more unbound material than the equal-mass binary while also forming a more massive post-collapse accretion disk. A particularly informative comparison can be made with the LS220 simulations of Lehner et al., which employ similar EOS and microphysics but neglect magnetic fields. For the equal-mass configuration, our GRMHD simulation produces approximately twice the ejecta mass reported in their hydrodynamic calculation, whereas the difference is considerably smaller for the unequal-mass case, where the tidal disruption effect dominates the mass ejection. Although differences in the binary parameters, neutrino treatment, and evolution time prevent a direct quantitative comparison, this trend suggests that magnetic stresses contribute significantly to the ejecta, particularly in nearly equal-mass mergers.

Our results further suggest that the ejecta continue to evolve well after the initial dynamical phase. Ciolfi \& Kalinani~(2020)~\cite{Ciolfi-Kalinani-2020} showed that long-lived magnetar remnants undergo a transition from an early dynamical ejecta phase to a delayed magnetically driven wind, which becomes the dominant source of mass ejection after several tens of milliseconds. On the other hand, Nedora et al.~(2021)~\cite{Nedora-2021} demonstrated that sustained postmerger mass loss can also be driven by spiral waves launched by the remnant. 
Although our HMNS collapses before such long-lived magnetically- and spiral-wave-driven winds can be established, the cumulative outflow mass continues to increase after merger, and the outflow rate exhibits a modest secondary enhancement, particularly in the unequal-mass model. Together with the stronger ($l=2,m=1$) gravitational wave mode observed in this case, these results suggest that the onset of a spiral-wave-driven outflow may already be present before collapse. In our GRMHD simulations, however, magnetic stresses are also expected to contribute to the continued postmerger mass loss, indicating that the observed secondary enhancement likely results from the combined action of non-axisymmetric spiral dynamics and magnetic stresses before BH formation.

\section{Conclusions}
\label{sec:conclusions}

In this work, we have presented the first binary neutron-star merger simulations performed with the \textsc{Spritz} code employing a finite-temperature tabulated equation of state. We considered two magnetized binaries with the LS220 equation of state and a GW170817-like chirp mass, differing only in their mass ratios ($q=1$ and $q=0.7$). Both systems were initialized with a strong purely poloidal magnetic field confined inside the neutron stars, while the electron fraction was passively advected and neutrino cooling was neglected. The simulations followed the binaries from the late inspiral through merger, the formation of a HMNS, and its subsequent collapse to a BH surrounded by an accretion disk. This work demonstrates the capability of \textsc{Spritz} to perform fully GRMHD simulations of BNS mergers with realistic finite-temperature microphysics. Moreover, it provides an investigation of how the binary mass ratio influences the postmerger evolution.

The equal-mass remnant survives for approximately 31.6\,ms after merger, whereas the unequal-mass remnant collapses after about 20.5\,ms, consistent with previous studies employing comparable binary configurations and the LS220 equation of state. The GW emission is dominated by the quadrupolar $(2,2)$ mode, with postmerger peak frequencies near 3\,kHz in both models. Both binaries also exhibit a clear post-merger $(l,m)=(2,1)$ mode associated with the development of non-axisymmetric structures in the HMNS, including the one-armed $m=1$ spiral instability. The mode is more pronounced in the unequal-mass model due to the initial asymmetry of the binary configuration.

The two binaries exhibit significantly different magnetic field evolution following the merger. In both models, the magnetic field is rapidly amplified immediately after the merger through KHI. Thereafter, the longer-lived equal-mass HMNS provides a more extended period during which differential rotation continuously winds up the poloidal magnetic field into a strong toroidal component, allowing the magnetic field to be further amplified prior to collapse. Consequently, the equal-mass model accumulates larger electromagnetic energies and poloidal magnetic fluxes and reaches isotropic-equivalent Poynting-flux luminosities approaching $10^{51}\,\mathrm{erg/s}$. Shortly before BH formation, we also observed the emergence of a collimated magnetically driven polar outflow, providing evidence for the development of an incipient jet during the HMNS phase. Following collapse, although the electromagnetic energy and magnetic flux gradually decline, the remnant develops a coherent magnetically dominated funnel, providing favorable conditions for the possible activation of the Blandford--Znajek mechanism. In contrast, the shorter-lived unequal-mass remnant undergoes a less extended phase of magnetic winding, resulting in weaker magnetic-field amplification and an earlier decline of both the electromagnetic energy and the poloidal magnetic flux. Consequently, it shows no clear evidence for a comparable collimated polar outflow or a highly magnetized funnel, despite being evolved for approximately twice as long after BH formation.

The binary mass ratio also has a significant impact on the properties of the postmerger ejecta. Due to the stronger tidal disruption during merger, the unequal-mass binary ejects approximately 50\% more mass than the equal-mass model and forms a more massive post-collapse accretion disk. As a result, it provides a larger reservoir of neutron-rich material for powering kilonova emission and possible late-time disk outflows. The enhanced mass ejection is consistent with the stronger non-axisymmetric dynamics of the remnant and the combined action of tidal disruption, spiral-wave, and magnetically driven mechanisms throughout the post-merger evolution. In contrast, the equal-mass model ejects less material while simultaneously developing conditions more favorable for magnetic field amplification and the emergence of an incipient jet. In this model, part of the baryonic material within the polar funnel is magnetically accelerated to velocities of approximately 0.2\,c and becomes gravitationally unbound already at relatively small distances from the remnant, suggesting the onset of a magnetically driven polar outflow prior to BH formation. These results demonstrate that the binary configuration most favorable for developing a magnetically driven incipient jet is not necessarily the one that launches the largest amount of neutron-rich ejecta. Instead, our simulations highlight the competing roles of binary mass ratio in shaping both the central engine responsible for short GRBs and the neutron-rich ejecta powering kilonova emission.

The conclusions presented here should be interpreted within the scope of the adopted physical and numerical assumptions. Although we applied a realistic microphysical equation of state, neutrino cooling and absorption are neglected, implying that the long-term thermodynamic evolution of the remnant cannot be fully captured. Furthermore, our simulations follow the post-collapse evolution for only a short time, preventing us from determining whether the formation of the highly magnetized funnel and observed polar outflow can ultimately lead to a successful Blandford-Znajek jet powered by the BH-disk system. Finally, the constant-density artificial atmosphere employed in the present calculations becomes increasingly important as the ejecta propagate to larger distances and would influence the subsequent jet-ejecta interaction. These limitations do not affect our principal conclusions regarding the early post-merger evolution but motivate more comprehensive simulations including improved microphysics, longer evolution times, and more realistic atmosphere treatments.

In summary, rather than identifying a single mechanism controlling the post-merger evolution of BNS mergers, our results reinforce the idea that the observable electromagnetic counterparts arise from the interplay of several competing physical processes. Our simulations reveal signatures of multiple mechanisms, including tidal disruption, KHI, magnetic winding, MRIs, non-axisymmetric spiral modes, and magnetically driven polar outflows, all operating simultaneously and influencing the remnant on different time scales. The relative importance of these processes depends sensitively on the binary properties, particularly the mass ratio, which governs the evolution of the system during merger and postmerger phases. Our study therefore highlights the importance of systematic GRMHD simulations spanning a broader range of binary parameters, equations of state, initial magnetic-field configurations, and microphysical ingredients. Only through such comprehensive studies will it become possible to quantify the relative roles of these mechanisms and establish robust connections between the properties of BNS mergers and the diverse electromagnetic counterparts observed in the era of multi-messenger astronomy.

\ack{We thank Domenico Logoteta and Rahime Matur for valuable discussions on the construction of the initial data and related aspects. This work is supported by the European Union under NextGenerationEU via the PRIN 2022 Project ``EMERGE'', Prot. n. 2022KX2Z3B (CUP C53D23001150006). 
JVK gratefully acknowledges support from the NSF grant OAC-2411068.
RC acknowledges additional support from the INAF Theory Grant 2023 ``AfterJet'', Ob.Fu. 1.05.23.06.02 (CUP C93C23006800005).
Numerical calculations have been made possible through a CINECA-INFN agreement and an award under the ISCRA initiative (Grant IsB32\_MerJet), both providing access to resources on LEONARDO at CINECA.}






\bibliographystyle{iopart-num}
\bibliography{references.bib}

\appendix
\section{Resolution study}
\label{sec:resolution_study}

\subsection{Assessment of MRI resolution}
\label{sec:MRI}

The MRI is expected to play a significant role in the amplification of magnetic fields and in the transport of angular momentum during the post-merger evolution of the remnant. To assess whether the fastest-growing MRI mode is adequately resolved in our simulations, we evaluate the MRI quality factor, $Q_{\mathrm{MRI}}= \lambda_{\mathrm{MRI}}/dx$, where $\lambda_{\mathrm{MRI}}$ is the wavelength of the MRI fastest-growing mode and $dx$ is the local grid spacing. We estimate the fastest-growing wavelength as $\lambda_{\mathrm{MRI}} = \frac{\sqrt{\pi}B}{|\Omega|\sqrt{\rho}}$, where $B$ is the magnetic field strength, $\Omega$ is the angular velocity, and $\rho$ is the rest-mass density. Following previous GRMHD studies, values of $Q_{\rm MRI} > 10$ are generally considered a necessary condition to resolve the fastest-growing MRI mode.

\begin{figure}
    \centering
    \includegraphics[width=0.9\linewidth]{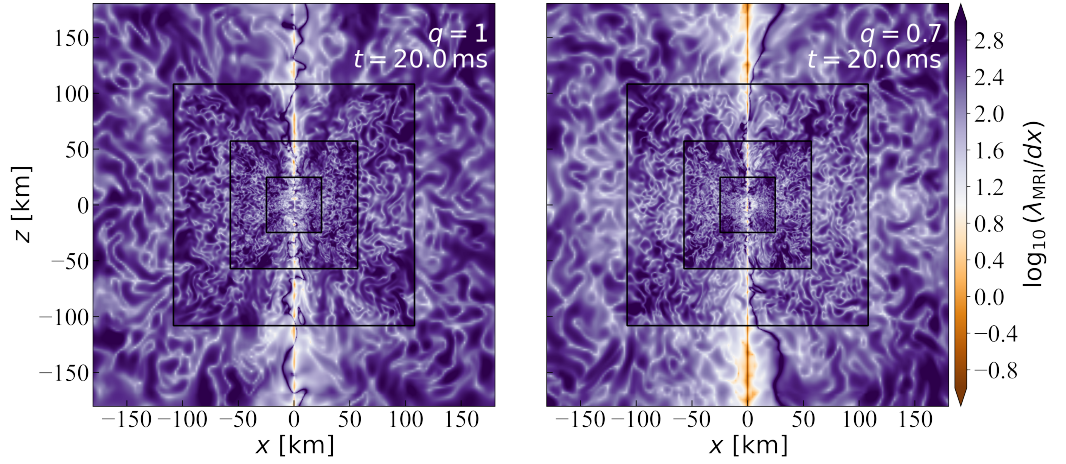}
    \caption{\justifying The MRI fastest growing mode wavelength divided by the grid spacing computed at $t=20$ms after the merger for each case.}
    \label{fig:LambdaMRI}
\end{figure}

Figure~\ref{fig:LambdaMRI} shows the 2D meridional view ($xz$-plane) of this quantity for the equal-mass and unequal-mass models approximately 20\,ms after merger. The nested boxes indicate the different refinement levels of our computational grid. 
In both simulations, the resolution criterion $Q_{\rm MRI} > 10$ is satisfied throughout most of the central remnant and the surrounding accretion flow.
Although the present resolution appears sufficient to capture the fastest-growing mode, it remains substantially lower than the ultra-high-resolution simulations in the literature, such as Kiuchi et al.~(2024)~\cite{Kiuchi-2024-alpha}, which resolve a much broader range of turbulent scales and demonstrate the operation of an MRI-driven $\alpha \Omega$ dynamo.

\subsection{Numerical convergence}
\label{sec:convergence}

To demonstrate the numerical robustness of the magnetic field evolution, we compare the equal-mass simulation performed with two grid resolutions, $dx=0.18 \,M_{\odot} (\simeq 266\,\rm{m})$ and $dx=0.15\,M_{\odot} (\simeq 222\,\rm{m})$. 

\begin{figure}[!htbp]
    \centering
    \includegraphics[width=0.98\linewidth]{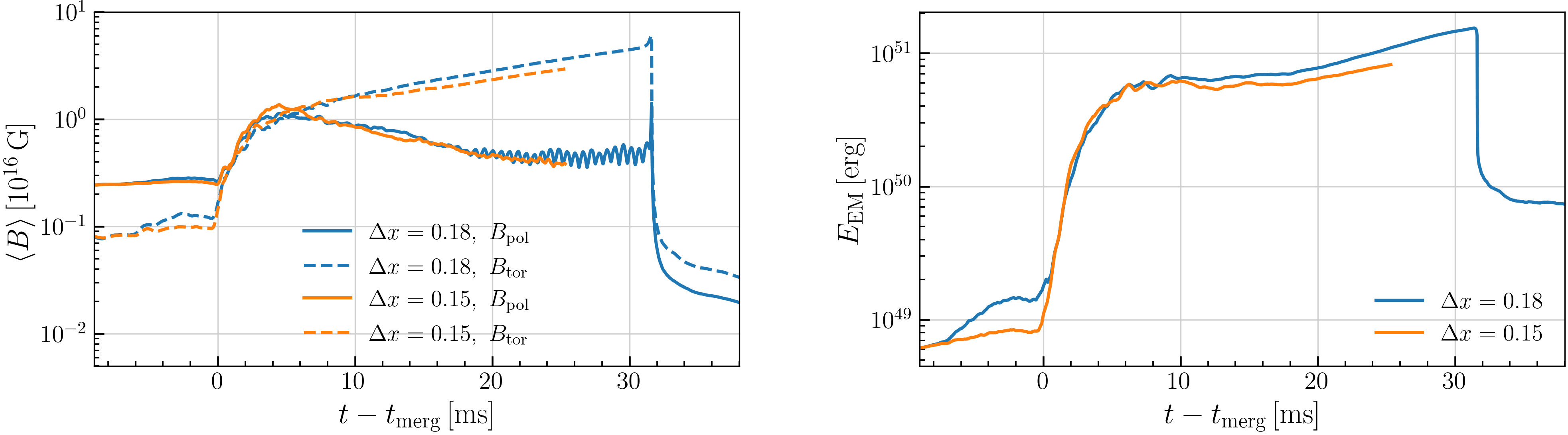}
    \caption{\justifying The resolution effects on magnetic field evolution for the equal-mass model; comparison for density-weighted volume-averaged toroidal and poloidal components (left), and the total magnetic energy (right).}
    \label{fig:resolution}
\end{figure}

The left panel of Fig.~\ref{fig:resolution} compares the evolution of the density-weighted volume-averaged poloidal and toroidal magnetic field components. The two simulations exhibit excellent agreement in the overall evolution of both components. In particular, the toroidal magnetic field grows continuously throughout the HMNS phase in both runs, confirming that magnetic winding is captured consistently. The poloidal component follows the same trend, reaching its maximum shortly after merger before gradually declining toward collapse. However, the oscillations observed during the late HMNS phase become noticeably weaker at higher resolution, indicating that they originate from numerical effects rather than representing a physical instability associated with the collapsing HMNS.
The right panel compares the evolution of the total electromagnetic energy (Eq.~\ref{eq:E_em}) for the two resolutions. The overall evolution is remarkably similar, with both simulations exhibiting rapid electromagnetic energy amplification following merger and continued growth throughout the HMNS phase. 

Overall, the higher-resolution simulation confirms the robustness of the magnetic field evolution presented in the main text. Increasing the resolution primarily suppresses numerical oscillations in the poloidal magnetic field while preserving the global evolution of the toroidal field and the total electromagnetic energy. We therefore conclude that the qualitative physical interpretation presented in this work is not affected by the adopted numerical resolution.

\end{document}